# Acoustic dispersion in a two-dimensional dipole system

Kenneth I Golden
Department of Mathematics and Statistics
College of Engineering and Mathematical Sciences
University of Vermont, Burlington, Vermont 05401-1455

Gabor J. Kalman
Department of Physics
Boston College, Chestnut Hill, Massachusetts 02467

Zoltan Donko and Peter Hartmann
Research Institute for Solid State Physics and Optics
of the Hungarian Academy of Sciences
H-1525 Budapest, P. O. Box 49, Hungary

We calculate the full density response function, and from it the long-wavelength acoustic dispersion for a two-dimensional system of strongly coupled point dipoles interacting through a $1/r^3$ potential at arbitrary degeneracy. Such a system has no RPA limit and the calculation has to include correlations from the outset. We follow the Quasi-Localized Charge (QLC) approach, accompanied by Molecular Dynamics (MD) simulations. Similarly to what has been recently reported for the closely spaced classical electron-hole bilayer [G. J. Kalman *et al.* Phys. Rev. Lett. **98**, 236801 (2007)] and in marked contrast to the RPA, we report a long-wavelength acoustic phase velocity that is wholly maintained by particle correlations and varies linearly with the dipole moment *p*. The oscillation frequency, calculated both in an extended QLC approximation and in the Singwi-Tosi-Land-Sjolander approximation, is invariant in form over the entire classical to quantum domains all the way down to zero temperature. Based on our classical MD-generated pair distribution function data and on ground-state energy data generated by recent quantum Monte Carlo simulations on a bosonic dipole system [Astrakharchik *et al*, Phys. Rev. Lett. **98**, 060405 (2007)], there is a good agreement between the QLCA kinetic sound speeds and the standard thermodynamic sound speeds in both the classical and quantum domains.

# I. INTRODUCTION

The formation of bound electron-hole excitons in semiconductors was predicted a long time ago by Keldysh *et al* and by Halperin and Rice [1]. Electron-hole bilayers (EHBs) have created an especially promising medium for the formation of stable excitons [2]. In such systems the charges in the two layers have opposite polarities, and for sufficiently small layer separations, the positive and negative charges bind to each other in dipole-like excitonic formations. Recent Monte Carlo (MC) studies [3, 4] have confirmed the emergence of the excitonic phase both in degenerate electron-hole [3] and in classical bipolar bilayers [4]. The simulations have shown the existence of four phases in the strong coupling regime: Coulombic liquid and solid and dipole liquid and solid phases. The necessary existence of these phases was also pointed out in Ref. [5]. In electron-hole bilayers, the excitons may also form a Bose-Einstein condensate [3, 6-8] or possibly a supersolid [5].

In a good approximation, the closely spaced EHB can be modeled as a two-dimensional (2D) monolayer of interacting point dipoles, each of mass $m = m_e + m_h$. The $N$ point dipoles are free to move in the $xy-$plane with dipolar moment oriented in the $z-$direction; the interaction potential is accordingly given by $\phi_D(r) = p^2/r^3$, where $p$ is the electric dipole strength. The approximation that replaces the bound electron-hole exciton by a point dipole has been considered by a number of investigators [9(a), 9(b), 10, 11].

The coupling strength in the EHB is characterized at arbitrary degeneracy by $\tilde{\Gamma} = e^2/a <E_{kin}>$, where $a = 1/\sqrt{\pi n}$ is the average in-plane distance between particles. In the high-temperature classical domain, this becomes the customary coupling parameter $\Gamma = \beta e^2/a$; $\beta = 1/k_B T$, while at zero temperature, it becomes $r_s = a/a_B$; $a_B = \hbar/m_{e,h} e^2$. By the same token, for the dipole system at arbitrary degeneracy, $\tilde{\Gamma}_D = p^2/a^3 <E_{kin}>$ can be taken as the coupling parameter, which becomes $\Gamma_D = \beta p^2/a^3$ in the high-temperature classical domain. At zero temperature, $\tilde{\Gamma}_D = r_D = r_0/a$ is the appropriate measure of the coupling strength; $r_0 = mp^2/\hbar^2$ is a characteristic length [9(a)]. Here we focus on the strong coupling regime $\tilde{\Gamma}_D >> 1$ that includes both the dipole liquid/solid phases. Since, in the symmetric ($m_e = m_h = m/2$) EHB, the Coulombic and dipole coupling parameters are related to each other by $r_D = 2(d^2/a^2)r_s$, high coupling ($r_D >> 1$) for point dipoles corresponds to the low-density regime in the closely spaced EHB, as dictated by the ordering $a > d >> a_B$ [3].

Various criteria have been put forward to determine whether the EHB can be considered as consisting of bound dipoles (excitons) rather than individual electrons and holes. In Ref. [3] the specific features of the correlation function showing a "correlation moat" surrounding the dipole was taken as the signature of the formation of permanent dipoles. Ref. [4] defined the dipole phase on the basis of comparing the energy of the assumed excitonic phase with that of a system of independent particles. In this paper, we will show that the 2D point-dipole model as a representation of the system of dipoles of finite size is justified on the basis of the comparison of the values of the average potential in the point dipole system with that in the EHB. All these criteria combined show that medium range $d/a$ (say $d/a \cong 0.6$) values are sufficiently low to maintain the validity of the point-dipole model. Thus, the requirement for exceedingly high $r_s$ values in the bilayer that seems to be required to balance closer layer separations (and which may be difficult to realize experimentally), can be avoided.

A variety of collective modes can exist in the strongly coupled EHB system [12, 13]. In the dipole approximation, the in-phase longitudinal mode can be identified with the density oscillation of the system of point dipoles. Prompted by this observation, there has been a recent flurry of activities directed at understanding the behavior of this collective mode both as a bilayer excitation and as a collective mode of the dipole system. A number of issues have emerged where results obtained through different approaches are at variance with each other. First, there is the central question of the dependence of the collective mode frequency on in-plane wavenumber *q*. In Ref. [5] an $\omega(q \to 0) \propto q^{3/2}$ dispersion has been proposed in the description of a Wigner (super)solid phase of dipoles. This behavior, in fact, can be regarded as the outcome of an application of the customary RPA argument to a dipole system: the lack of validity of this approach is pointed out below. By contrast, all the Ref. [9, 10, 13] investigators report an $\omega(q \to 0) \propto q$ acoustic dispersion, albeit arrived at through different approaches. A second issue is the dependence of $\omega(q \to 0)$ on the layer separation distance *d*: the RPA analysis of the dipole system in Ref. [10] asserts that $\omega(q \to 0) \propto \sqrt{d}$, whereas the analysis of Ref. [13], which takes account of correlational effects from the outset, indicates that $\omega(q \to 0) \propto d$; so do the Refs. [9] studies of strongly correlated dipoles. Finally, there is the question of the precise numerical value of the phase velocity of the mode for which different values have been put forward by Refs. [9, 10, 13].

In this paper, we will rigorously show that the actual dispersion for the density oscillations in the dipole liquid is acoustic, i.e., $\omega(q \to 0) \sim q$ and that it can be attributed to dipole dynamics, i.e., $\omega(q \to 0) \sim d$ ($d$ being proportional to the dipole moment $p$); we will elucidate the correspondence between this result and a similar finding for the in-phase mode in the strongly correlated EHB.

One would expect that the RPA is an appropriate approach at least for the qualitative description of the collective modes both in the EHB and in the dipole systems. In fact, this expectation has not been borne out. The application of the RPA to an EHB leads to an acoustic $q \to 0$ behavior for the in-phase mode, but with an acoustic velocity $s = \omega_0 \sqrt{ad}$, where $\omega_0^2 = 4e^2/(ma^3)$ is the nominal 2D plasma frequency. This is not surprising, since the inappropriateness of the RPA for the analysis of the EHB has already been pointed out in [13]: the RPA treatment cannot reproduce the merging of the intrinsic dipole oscillation with the out-of-phase collective mode. For the analysis of the dipole liquid, the inadequacy of the RPA has a different origin and is more grievous. An attempt to establish an RPA formalism fails entirely because the average Hartree field

$$<\phi_D(r)>_H = n \int d^2\mathbf{r} \phi_D(r)$$

of the dipole potential $\phi_D(r) = p^2/r^3$ diverges. Therefore, the Fourier transform of the dipole potential does not exist, implying that the 2D system of point dipoles interacting via this potential can have no RPA limit. As a consequence, the routine argument that would generate the RPA collective mode frequency via

$$\omega^2(q) \propto q^2 \phi_D(q); \quad \phi_D(q) = -q^2 \phi_{COUL}(q) d^2$$

($\phi_{COUL}(q) = 2\pi e^2/q$), [5] becomes invalid, ruling out the ensuing $\omega(q \to 0) \propto q^{3/2}$ dispersion.

A correlational non-RPA study of the collective mode spectrum of the EHB *liquid* in the classical domain was carried out in the recent work by the authors [13] through a combined analytical/molecular dynamics (MD) approach. The analytical portion of the study was based on the quasilocalized charge approximation (QLCA) [14], which in the $d \to 0$ dipole limit led to a long-wavelength acoustic dispersion with phase velocity

$$s(d \to 0) = \sqrt{\frac{2K}{m} <\phi_D(r)>} = \omega_0 d \sqrt{KI} \;. \tag{1}$$

This relationship was also corroborated by the accompanying MD simulations [14]. With $g_{11}(\bar{r})$, the inlayer pair distribution function, the average potential $<\phi_D(r)>$ is defined as

$$<\phi_D(r)> = n \int d^2\mathbf{r} \, g_{11}(r) \phi_D(r) = 2(p^2/a^3)I, \quad I = I(\Gamma, d) = \int_0^\infty d\bar{r} \, g_{11}(\bar{r}) \frac{1}{\bar{r}^2},$$

$\bar{r} = r/a$; $K = 33/32$ as calculated for the QLCA. For $d/a = 0.6$, the integral $I(\Gamma, d)$ is of the order of 0.8: its precise value and its dependence on $\Gamma$ and $d$ will be discussed in Section V. The expression (1) is now in marked contrast to the RPA acoustic velocity $s = \omega_0 \sqrt{ad}$.

The analysis of Ref. [13] was extended to the solid phase of the EHB by applying the conventional harmonic approximation for phonons and summing over lattice sites: the dispersion of the longitudinal phonon in the $d \to 0$ dipole limit at $T = 0$ is given by a formula similar to Eq. (1), with a slightly different value of the integral; details will be given in Section III.

A similar calculation for the EHB lattice phonons was carried out by Kulakovskii *et al* [12]. These authors also derived an acoustic dispersion, but with a coefficient at variance with Eq. (1): $s(d \to 0) = \pi^{1/4} \omega_0 d$.

Turning now tho the 2D point-dipole system, Kachintsev and Ulloa [10] were the first to analyze the collective excitations in a 2D fluid of bosonic dipoles modeled as point dipoles. They introduced a softened interaction potential $\phi_{KU}(r) = \phi_D(r)[1 - \exp(-r^2/d^2)]$; such a modification, which is Fourier-transformable, makes it possible to realize an average Hartree field $n \int d^2\mathbf{r} \phi_{KU}(r)$ and, consequently, an RPA limit resulting in an acoustic $\omega(q \to 0) \propto q$ dispersion with acoustic velocity $s = 0.15 \omega_0 \sqrt{ad}$.

Convincing evidence for the acoustic behavior in the point-dipole system comes from recent quantum Monte Carlo (QMC) simulations of the degenerate bosonic dipole fluid in high [9(a)] and rather weak to low [9(b)] coupling regimes. These data provide indirect evidence for the acoustic dispersion through the application of the Feynman relation [see Eq. (45) below] with the input of computed static structure function $S(q)$ data. More will be said in Sec. V about quantitative comparisons of theory with the Ref. [9(a)] simulations in the strong coupling regime.

The QMC simulation of the degenerate bosonic dipole system was extended beyond the freezing point into the lattice phase [9(a)]. The results are not qualitatively different from those for the liquid phase and will be discussed in Section III.

After this lengthy preamble, we can state in precise terms the purpose of this paper: it is to approach the question of the small-$q$ dispersion of the strongly coupled excitonic fluid by studying directly the collective mode dispersion of the strongly coupled 2D *dipolar liquid*. We will show that the dispersion is acoustic, and we will study its characteristic sound velocity and its relationship to the corresponding quantity in the EHB. This analysis is the central objective of the present work.

We propose to calculate the collective mode behavior by invoking two well-tested and rather different approaches: (i) the QLCA dynamical equation-of-motion/collective coordinates approach [14] and (ii) the Singwi-Tosi-Land-Sjölander (STLS) kinetic equation approach [15–17]. There is no need to explicitly specify the degree of degeneracy in either formalism (Secs. II, VI). However, the more involved STLS analysis (Section VI) is carried out first in the high-temperature classical domain and then in the quantum domain at arbitrary temperatures. We contend that in the strong coupling regime the QLCA is superior to the STLS approach: this is borne out by the comparison of the model theoretical results both with standard thermodynamic results and with those generated from computer simulations. Nevertheless, we follow this strategy in order to illustrate that in the domain of interest, quantum effects have no bearing on the architecture of the collective mode dispersion. There remains one open question, namely whether the formation of a condensate would affect the mode dispersion. That this indeed may be the case is known from the Bogoliubov analysis of the excitation spectrum of weakly interacting bosons. Here, however, the Bose-Einstein condensate fraction can be considered to be negligibly small since strong dipole-dipole interactions tend to destroy coherence. This observation is borne out by the Ref. [9(a)] QMC simulation.

The essential point to be noted in all the theoretical calculations is that, as discussed above, the dipole potential does not admit an RPA-like approximation. Hence, the density response function for the $1/r^3$ interaction can be calculated only through the introduction of correlations in the formalism from the outset. This is why it is crucial to rely on a non-perturbative calculational method, such as the proposed QLCA and STLS theoretical approaches.

Our theoretical analysis is accompanied by a Molecular Dynamics study of the strongly coupled classical dipole liquid. While the full scope of this work will be reported elsewhere [22], here we will cite the conclusions that pertain to the low-$q$ behavior of the density oscillation mode.

As to the plan of the rest of the paper, In Sec. II, we reformulate the QLCA, which was originally created for the classical charged liquid [14], into an approximation method suitable for the description of the 2D system of strongly interacting point dipoles. This will be done through a two-stage procedure: (i) first develop a classical QLCA theory along the lines of Ref. [14] to be followed by (ii) its extension into the quantum domain along the lines of Ref. [18]. In Section III, we use the harmonic approximation to calculate the long-wavelength acoustic behavior of the longitudinal phonon in the 2D dipole crystal; we also compare the results with those of a similar calculation for the phonons in the EHB crystal [13]. In Sec. IV, we calculate thermodynamic sound speeds of the 2D dipole liquid both in the classical and in the zero-temperature quantum domains. These may serve as standards for comparison in Sec. V where we establish linkages between the Sec. II QLCA sound velocity and the corresponding data generated from our classical MD and the Ref. [9(a)] quantum MC simulations; we also provide comparisons with the results of our earlier work [13] on the sound velocity in the EHB. In Sec. VI, in order to see how quantum effects may or may not alter the semi-classical results, we adapt the classical and quantum STLS kinetic equation descriptions of the 3D electron gas [15, 16] (using the more tractable Ref. [19] quantum kinetic equation formalism) to the strongly coupled 2D point-dipole liquid. Conclusions are drawn in Sec. VII.

## II.  QLCA DESCRIPTION

We turn now to the formulation of a QLC approximation scheme for the model 2D monolayer of $N$ strongly interacting point dipoles. Let $A$ be the large but bounded area of the monolayer and $n = N/A$ the average density. In the two-stage development of the extended QLCA, we begin with the derivation in the classical domain.

The QLC method has already been established and successfully applied to strongly coupled charged particle systems. Here we follow the paradigm of the original derivation, focusing on the differences that distinguish the point dipole system from a system of point charged particles. Similarly to what has been established for charged particle systems, the observation that serves as the basis of the QLC theory is that the dominating feature of the physical state of a classical dipolar liquid with coupling parameter $\Gamma_D \gg 1$ is the quasi-localization of the point dipoles. The ensuing model closely resembles a disordered solid where the dipoles occupy randomly located sites and undergo small-amplitude oscillations about them. However, the site positions also change and a continuous rearrangement of the underlying quasi-equilibrium configuration takes place. Inherent in the model is the assumption that the two time scales are well separated and that it is sufficient to consider the time average (converted into ensemble average) of the drifting quasi-equilibrium configuration.

In the first stage, we wish to calculate the linear response to a weak perturbing external dipole potential energy $\Phi_D^{ext}$. Following the procedure of [14], let $\mathbf{X}_i(t) = \mathbf{x}_i + \boldsymbol{\xi}_i(t)$ be the momentary position of the ith point dipole, $\mathbf{x}_i$ its quasi-equilibrium site position, and $\xi_i(t)$ the perturbed amplitude of its small excursion; $\xi_i(\omega)$ is its Fourier transform. In the equations that follow, $i$, $j$ subscripts enumerate particles and $\mu, \nu$ are vector indices; Einstein summation convention of the repeated vector indices is understood. The microscopic equation of motion for the ith dipole is

$$-m\omega^2 \xi_{i,\mu}(\omega) + \sum_j K_{ij,\mu\nu} \xi_{j,\nu}(\omega) = -\frac{\partial}{\partial x_{i,\mu}} \Phi_D^{ext}(\mathbf{x}_{i,\mu}, \omega), \qquad (2)$$

$$K_{ij,\mu\nu} = (1-\delta_{ij})\frac{\partial^2 \phi_{ij}}{\partial x_{i,\mu} \partial x_{j,\nu}} - \delta_{ij} \sum_\ell (1-\delta_{i\ell})\frac{\partial^2 \phi_{i\ell}}{\partial x_{i,\mu} \partial x_{\ell,\nu}}; \qquad (3)$$

$\phi_{ij} = p^2 / |\mathbf{x}_i - \mathbf{x}_j|^3$ is the point dipole potential. Eq. (3) shows the characteristic separation of the potential energy $(1/2)\sum_{i,j} K_{ij,\mu\nu} \xi_{i,\mu} \xi_{j,\nu}$ into diagonal ($\delta_{ij}$) and off-diagonal $[(1-\delta_{ij})]$ contributions: the former originates from the displacement of a dipole in a fixed environment of the other dipoles, while the latter originates from the fluctuating environment.

We next introduce collective coordinates $\xi_\mathbf{k}$ via the Fourier representation

$$\xi_{i,\mu}(\omega) = \frac{1}{\sqrt{mN}} \sum_\mathbf{k} \xi_{\mathbf{k},\mu}(\omega) \exp(i\mathbf{k} \cdot \mathbf{x}_i). \qquad (4)$$

Substituting Eq. (4) into (2) and following the procedure of [14], one ultimately obtains the ensemble-averaged equation of motion in terms of the dynamical tensor $C_{\mu\nu}(\mathbf{q})$:

$$\left[\omega^2 \delta_{\mu\nu} - C_{\mu\nu}(\mathbf{q})\right] \xi_{\mathbf{q},\nu}(\omega) = \frac{inq_\mu}{\sqrt{mN}} \Phi_D^{ext}(\mathbf{q}, \omega), \qquad (5)$$

$$C_{\mu\nu}(\mathbf{q}) = \frac{1}{mN}\sum_{i,j} < K_{ij,\mu\nu} \exp[-i\mathbf{q}\cdot(\mathbf{x}_i - \mathbf{x}_j)] >$$

$$= \frac{3np^2}{m}\int d^2\mathbf{r}\,\frac{1}{r^5}g(r)[\exp(i\mathbf{q}\cdot\mathbf{r})-1]\left[\delta_{\mu\nu} - 5\frac{r_\mu r_\nu}{r^2}\right], \tag{6}$$

where $g(r)$ is the equilibrium pair distribution function. Projecting out the longitudinal (LL: with respect to **q**) element of the dynamical tensor, we derive an equation for the average density response $n(\mathbf{q},\omega)$ by using the relation $n(\mathbf{q},\omega) = -(iqN/\sqrt{mN})\xi_{\mathbf{q},L}(\omega)$:

$$\left[\omega^2 - C(\mathbf{q})\right]n(\mathbf{q},\omega) = \frac{nq^2}{m}\Phi_D^{ext}(\mathbf{q},\omega), \tag{7}$$

$$C(\mathbf{q}) \equiv C_{LL}(\mathbf{q}) = \frac{3\pi np^2}{m}\int_0^\infty dr\,\frac{1}{r^4}g(r)\left[3 - 3J_0(qr) + 5J_2(qr)\right], \tag{8}$$

It is useful to introduce the notation

$$\Psi(\mathbf{q}) = \frac{m}{nq^2}C(\mathbf{q}).$$

Eq. (7) and the constitutive relation

$$n(\mathbf{q},\omega) = \chi(\mathbf{q},\omega)\Phi_D^{ext}(\mathbf{q},\omega) \tag{9}$$

then give the QLCA density response function

$$\chi(\mathbf{q},\omega) = \frac{nq^2/m\omega^2}{1 - \Psi(q)nq^2/m\omega^2}, \tag{10}$$

$$\Psi(q) = \frac{3\pi p^2}{q^2}\int_0^\infty dr\,\frac{1}{r^4}g(r)\left[3 - 3J_0(qr) + 5J_2(qr)\right]. \tag{11}$$

Note that the usual procedure of splitting $\Psi(q)$ into RPA and correlational parts by replacing $g(r)$ by $1+h(r)$ does not work, because both of the separated terms would be represented by divergent integrals.

We observe that the derivation of (10) is predicated on the reasonable assumption that thermal motions are negligible in the high coupling regime. Nevertheless, the effects of random motion of the particles can be incorporated in the formalism [20] by replacing the $nq^2/m\omega^2$ factors in (10) by the Vlasov density response function $\chi_0^V(\mathbf{q},\omega)$ (in the classical domain) or by the Lindhard density response function $\chi_0^L(\mathbf{q},\omega)$ (in the quantum domain).

In the quantum domain where the fluctuations are more important, the second-stage reformulation of the QLCA parallels the procedure of Ref. [18]. One may accordingly assume that the effect of random quantum fluctuations is well represented by replacing the $nq^2/m\omega^2$ factors by the Lindhard function

$$\chi_0^L(\mathbf{q},\omega) = \frac{1}{A} \sum_{\mathbf{k}} \left[ \frac{n_{\mathbf{k}-\mathbf{q}/2} - n_{\mathbf{k}+\mathbf{q}/2}}{\hbar\omega - (\hbar^2/m)\mathbf{k}\cdot\mathbf{q}} \right], \qquad (12)$$

resulting in the density response function

$$\chi(\mathbf{q},\omega) = \frac{\chi_0^L(\mathbf{q},\omega)}{1 - \Psi(q)\chi_0^L(\mathbf{q},\omega)}, \qquad (13)$$

that replaces Eq. (10) for <u>arbitrary degeneracy</u>; $n_{\mathbf{k}}$ is the momentum distribution function for particles with energy spectrum $\varepsilon_{\mathbf{k}} = \hbar^2 k^2/(2m)$; $N = \sum_{\mathbf{k}} n_{\mathbf{k}}$ with $n = N/A$ the average 2D density. At strong coupling and in the $q \to 0$ limit, $\chi_0^L(q \to 0, \omega) \approx nq^2/m\omega^2$, so that (10) and (13) coincide and both the extended quantum QLC and classical QLC approximations lead to

$$\chi(q \to 0, \omega) = \frac{nq^2/m\omega^2}{1 - \Psi(q \to 0)nq^2/m\omega^2}, \qquad (14)$$

$$\Psi(q \to 0) = \frac{33}{8}\pi p^2 \int_0^\infty dr \frac{1}{r^2} g(r) = \frac{33}{16}\pi a^2 <\phi_D(r)>. \qquad (15)$$

Defining the dipole oscillation frequency (equivalent of the 2D plasma frequency)

$$\omega_D^2 = \frac{2\pi p^2 n}{ma^3},$$

the acoustic mode oscillation frequency then follows from setting the denominator of (14) equal to zero:

$$\omega^2(q \to 0) = \frac{33}{8}\frac{\pi n p^2}{m} q^2 \int_0^\infty dr \frac{1}{r^2} g(r) = \frac{33}{16} J(\tilde{\Gamma}_D)\omega_D^2 a^2 q^2 \qquad (16)$$

We thus obtain the phase velocity as

$$s = \omega_D a \sqrt{2KJ(\tilde{\Gamma}_D)} \qquad (17)$$

with $K = 33/32$ and

$$J(\tilde{\Gamma}_D) = \int_0^\infty d\bar{r} \frac{1}{\bar{r}^2} g(\bar{r}); \qquad (18)$$

$\bar{r} = r/a$. We note that $J(\tilde{\Gamma}_D)$ is identical to $I(\tilde{\Gamma}, d)$ in (1), except for the difference in the correlation functions under the integral: $I(\tilde{\Gamma}, d)$ is defined through the inlayer correlation function of the bilayer and has a weak dependence on $d$, while in $J(\tilde{\Gamma}_D)$ the correlation function is that of the point dipoles. Tabulated values of the $J(\tilde{\Gamma}_D)$ and $I(\Gamma,d)$ integrals are displayed in Tables 2 and 4, respectively.

To see that $g(r \to 0)$ tends to zero sufficiently fast to guarantee the convergence of the integral in Eqs. (15), (16) and (18), we observe that this is indeed the case in the high-temperature classical domain where one would expect that $g(r \to 0) \propto \exp(-\beta p^2 / r^3)$. In fact, this has been verified by our MD simulation. To make the case for convergence in the zero-temperature quantum domain, we observe that when two point dipoles are in close proximity to each other, the pair wave function $\psi(r)$, and consequently the pair distribution function $g(r) \propto |\psi(r)|^2$, are determined by the solution to the two-particle Schrödinger equation in the $r \to 0$ limit. Paralleling Kimball's electron gas calculation [21], one readily finds that

$$\psi(r \to 0) = K_0\left(2\sqrt{r_0/r}\right) \approx \frac{\sqrt{\pi}}{2}\left[\frac{r}{r_0}\right]^{1/4} \exp\left(-2\sqrt{r_0/r}\right), \tag{19}$$

with the characteristic length $r_0 = mp^2/\hbar^2$ introduced above; $K_0\left(2\sqrt{r_0/r}\right)$ is the modified Bessel function of the second kind. This small-$r$ behavior has also been reported in Ref. [9(a)]. Consequently,

$$g(r \to 0) \propto \frac{\pi}{4}\sqrt{\frac{r}{r_0}} \exp(-4\sqrt{r_0/r}), \tag{20}$$

again guaranteeing the convergence of the integral in Eqs. (15), (16), and (18).

## III. DIPOLE SOLID

The philosophy of the QLCA scheme not being substantially different from that of the harmonic approximation for lattice phonons, the results of the previous Section can be converted, *mutatis mutandis*, into a description of the $q \to 0$ dispersion of lattice phonons. Based on Eq. (5), the lattice dispersion relation can be written as

$$\|\omega^2 \delta_{\mu\nu} - C_{\mu\nu}(\mathbf{q})\| = 0, \tag{21}$$

$$C_{\mu\nu}(\mathbf{q}) = \frac{3np^2}{m} \sum_{i \neq 0} [\exp(i\mathbf{q} \cdot \mathbf{r}_i) - 1] \frac{1}{r_i^5} \left[ \delta_{\mu\nu} - 5\frac{r_{i\mu} r_{i\nu}}{r_i^2} \right] \tag{22}$$

Since to $O(q^2)$ the triangular lattice exhibits isotropic behavior, one can focus on the longitudinal dispersion and obtain to $O(q^2)$

$$\omega^2(q \to 0) = C(q \to 0) = \frac{33}{32} M \omega_D^2 a^2 q^2, \tag{23}$$

or

$$s_{SOLID} = \omega_D a \sqrt{\frac{33}{32} M}. \tag{24}$$

where

$$M = \sum_i \frac{1}{\bar{r}_i^3} \tag{25}$$

is the lattice sum over the triangular lattice; $\bar{r}_i \equiv r_i / a$. In effect, $M/2$ replaces the integral $J(\tilde{\Gamma}_D)$ in Eq. (16). The value of $M$ has been calculated by a number of workers [24-27] with slightly different results; the most recent semi-analytic calculation is due to Rozenbaum [27], according to which

$$\frac{M}{2} = \frac{1}{2}11.341\left[\frac{\sqrt{3}}{2\pi}\right]^{3/2} = 0.821 \tag{26}$$

Our own lattice sum computation for the 2D dipole crystal involving $1.9 \times 10^9$ particles provides

$$\frac{M}{2} = 0.7985; \tag{27}$$

From (24), the corresponding sound speed is then

$$s_{SOLID} = 1.283\omega_D a. \tag{28}$$

This can be compared with the quantum MC formula for the potential energy of the dipole crystal quoted from Ref. [9(a)] as

$$E_{triang} = 4.446(nr_0^2)^{3/2} E_0 = \frac{p^2}{2a^3} M; \tag{29}$$

$E_0$ is defined in Eq. (31) below. From (29) we calculate

$$\frac{M}{2} = \frac{4.446}{\pi^{3/2}} = 0.7984, \tag{30}$$

so that the sound speed (28) again results.

As to the linkage with the classical zero-temperature EHB crystal, our results (27) and (28) can be compared with our corresponding EHB lattice sums $L(d)$, which, of course, depend on the layer separation:

$$L(d) = \sum_i \left[\frac{1}{\bar{r}_i} - \frac{1}{\sqrt{\bar{r}_i^2 + \bar{d}^2}}\right] \tag{31}$$

The $L(d)$ values, together with their associated sound velocities, are tabulated in Table 1.

| $d/a$ | $L(d)$ | $s_{SOLID}$ $(a\omega_D)$ |
|---|---|---|
| 0.1 | 0.7974 | 1.282 |
| 0.2 | 0.7944 | 1.280 |
| 0.3 | 0.7895 | 1.276 |
| 0.4 | 0.7828 | 1.271 |
| 0.5 | 0.7744 | 1.264 |
| 0.6 | 0.7646 | 1.256 |

**Table 1.** EHB Lattice sum and sound velocity (in $a\omega_D$ units) as functions of layer separation $d$.

The tabulated $L(d)$ and sound speed values are quite close to the Eqs. (27) and (28) $M/2$ and sound speed values, even for $d/a = 0.6$ as expected.

## IV. DIPOLE LIQUID-PHASE THERMODYNAMICS

We turn next to the straightforward derivation of the thermodynamic sound speed in the 2D point-dipole liquid. We consider first the classical domain. Starting from the correlation energy per particle of the dipole system

$$E_{CORR} = \frac{n}{2}\int d^2\mathbf{r}\,\phi_d(r)g(r) = \frac{p^2}{a^3}J(\Gamma_D),  \qquad (32)$$

or more succinctly,

$$\beta E_{CORR} = \Gamma_D J(\Gamma_D);  \qquad (33)$$

the $J(\Gamma_D)$ integral is defined in Eq. (18) above. The total thermodynamic pressure $P$ is calculated to be

$$\frac{\beta P}{n} = 1 + \frac{3}{2}\Gamma_D J(\Gamma_D), \qquad (34)$$

and the dipole sound speed formula

$$\beta m s^2(\Gamma_D) = \beta \frac{\partial P}{\partial n} = 1 + \frac{15}{4}\Gamma_D J(\Gamma_D) + \frac{9}{4}\Gamma_D^2 J'(\Gamma_D) \qquad (35)$$

readily follows from (34). To facilitate comparison with the QLCA sound speed (17), Eq. (35) can also be recast in $a\omega_D$ units:

$$s = \omega_D a \sqrt{\frac{1}{2\Gamma_D} + \frac{15}{8} J(\Gamma_D) + \frac{9}{8}\Gamma_D J'(\Gamma_D)} \qquad (36)$$

Isothermal sound speed values generated from (36) are tabulated in Table 2 over a wide range of liquid-phase coupling strengths. We note that the QLCA values are a few percent *above* the corresponding thermodynamic values.

| $\Gamma_D$ | $J(\Gamma_D)$ | $s_{QLCA}$ ($a\omega_D$) | $s_{MD}$ ($a\omega_D$) | $s_{COMP}$ ($a\omega_D$) | $s_{MD}$ ($1/\sqrt{\beta m}$) | $s_{COMP}$ ($1/\sqrt{\beta m}$) | $s_{QLCA}$ ($1/\sqrt{\beta m}$) |
|---|---|---|---|---|---|---|---|
| 20 | 0.8865 | 1.352 | 1.339 | 1.272 | 8.47 | 8.042 | 8.55 |
| 40 | 0.8515 | 1.325 | 1.268 | 1.251 | 11.34 | 11.19 | 11.85 |
| 60 | 0.8367 | 1.314 | 1.281 | 1.243 | 14.03 | 13.62 | 14.39 |
| 80 | 0.8298 | 1.308 | 1.284 | 1.239 | 16.24 | 15.67 | 16.55 |
| 100 | 0.8245 | 1.304 | 1.285 | 1.235 | 18.17 | 17.47 | 18.44 |

**Table 2**. 2D-point diple liquid: QLCA ($s_{QLCA}$), MD ($s_{MD}$), and thermodynamic ($s_{COMP}$) sound speeds as functions of the classical coupling parameter $\Gamma_D$. Columns 3 - 5 are in units of $a\omega_D$; columns 6 - 8 are in units of $1/\sqrt{\beta m}$.

We turn next to the derivation of the thermodynamic sound speed in the 2D bosonic dipole liquid at zero temperature. The starting point for the calculation is the ground-state energy fitting formula given by the quantum MC simulation work [9(a)] in the strong coupling regime of the degenerate dipole liquid:

$$E = [a_1(nr_0^2)^{3/2} + a_2(nr_0^2)^{5/4} + a_3(nr_0^2)^{1/2}]E_0; \quad (4 \le nr_0^2 \le 256) \quad (37),$$

$E_0 = \hbar^2/mr_0^2$ is the dipole equivalent of the Rydberg energy; $r_0 = mp^2/\hbar^2$,

$a_1 = 4.536$, $a_2 = 4.38$, and $a_3 = 1.2$.

The thermodynamic pressure and sound speed are then calculated to be

$$P = n^2 \frac{\partial E}{\partial n} = n(nr_0^2)\left[\frac{3}{2}a_1(nr_0^2)^{1/2} + \frac{5}{4}a_2(nr_0^2)^{1/4} + \frac{1}{2}a_3(nr_0^2)^{-1/2}\right]E_0, \quad (38)$$

$$ms^2 = \frac{\partial P}{\partial n} = nr_0^2\left[\frac{15}{4}a_1(nr_0^2)^{1/2} + \frac{45}{16}a_2(nr_0^2)^{1/4} + \frac{3}{4}a_3(nr_0^2)^{-1/2}\right]E_0. \quad (39)$$

Or, in terms of the convenient $a\omega_D$ units, (35) becomes:

$$s^2 = \frac{a^2\omega_D^2}{2\pi^{3/2}}\left[\frac{15}{4}a_1 + \frac{45}{16}a_2(nr_0^2)^{-1/4} + \frac{3}{4}\frac{a_3}{nr_0^2}\right] \quad (4 \le nr_0^2 \le 256) \quad (40)$$

Sound speeds generated from (39) and (40) are tabulated in Table 3 over a wide range of liquid phase coupling strengths. We note that, in contrast to the classical case, the QLCA values are a few percent *below* the corresponding thermodynamic values.

| $nr_0^2$ | $r_D$ | $s_{QMC}$ $(a\omega_D)$ | $s_{QLCA}$ $(a\omega_D)$ | $s_{COMP}$ $(a\omega_D)$ | $s_{QLCA}$ $\left(\sqrt{E_0/m}\right)$ | $s_{COMP}$ $\left(\sqrt{E_0/m}\right)$ |
|---|---|---|---|---|---|---|
| 32 | 10.0 | ~1.76 | 1.3 | 1.41 | 58.2 | 63.3 |
| 64 | 14.2 | ~1.70 | 1.3 | 1.39 | 97.9 | 105 |
| 128 | 20.0 | ~1.69 | 1.3 | 1.36 | 164.7 | 172.7 |
| 256 | 28.4 | ~1.80 | 1.3 | 1.34 | 277.7 | 286.2 |

**Table 3.** 2D point-dipole liquid: QLCA ($s_{QLCA}$), quantum MC ($s_{QMC}$), and thermodynamic ($s_{COMP}$) sound speeds as a function of the zero-temperature coupling parameter $r_D = r_0/a$. Columns 3 - 5 are in units of $a\omega_D$. Columns 6 and 7 are in units of $\sqrt{E_0/m}$; $E_0 = \hbar^2/mr_0^2$

## V. SIMULATIONS

In the strong coupling regime, we expect that the collective mode behavior is well emulated by a classical model. In order to further study the collective mode behavior and to assess the validity of the QLCA, we have performed a classical MD simulation of the 2D dipole liquid. Details of this method and of the result for the full dispersion of the entire mode spectrum will be described elsewhere [22]. Here we quote the relevant $q \to 0$ results for the longitudinal collective mode. The values of the integral $J(\Gamma_D)$ as calculated from MD simulated pair distribution functions and the QCLA as well as MD sound velocities are given in Table 2. Using the thermodynamic sound velocity (36) as a reference, the discrepancy between QLCA and thermodynamic sound velocities ranges from 5.55% at $\Gamma_D = 100$ to 6.34% at $\Gamma_D = 20$. As expected, the MD sound velocity is somewhat closer to the thermodynamic sound velocity.

As to correspondence with the classical EHB liquid, we see from Tables 2 and 4 that the EHB and 2D point dipole respective $I(\Gamma,d)$ and $J(\Gamma_D)$ values are quite close. In general, the EHB sound speeds are lower than their point-dipole counterparts. For example, at $\Gamma = 121$ and $d/a = 0.6$, the EHB sound speed based on Eq. (1) is ~1.2% lower than that computed for the 2D point dipole liquid at the corresponding $\Gamma_D = 43.6$ coupling strength, but this is within the uncertainties of the determination of the sound speed from the MD data. We can also compare the point-dipole QLCA sound velocities (Table 2, column 3) with the "exact" [i.e., calculated without the assumption of linear dependence on $d$ in $\omega(q,d)$] EHB QLCA phase velocities (Table 4, column 5). Again the EHB sound speeds are lower, with a somewhat larger difference, ranging from 7.6% at $\Gamma_D = 20$ to 6.2% at $\Gamma_D = 100$.

| $\Gamma$ | $\Gamma_D$ | $I$ | $s_{QLCA}$ [Eq. (1)] | $s_{QLCA}$ [Ref. 13] |
|---|---|---|---|---|
| 52 | 18.72 | 0.8506 | 1.325 | 1.251 |
| 60 | 21.60 | 0.8449 | 1.320 | 1.247 |
| 105 | 37.80 | 0.8309 | 1.304 | 1.236 |
| 121 | 43.56 | 0.8285 | 1.307 | 1.234 |
| 212 | 76.32 | 0.8186 | 1.299 | 1.226 |
| 243 | 87.48 | 0.8140 | 1.296 | 1.224 |
| 280 | 100.8 | 0.8126 | 1.295 | 1.223 |

**Table 4.** EHB: Integral $I$ and QLCA sound speeds (in $a\omega_D$ units) as a function of $\Gamma$ for $d/a = 0.6$. The QLCA sound speeds in column 4 are calculated from Eq. (1) with the input of column 3. The more exact sound speed values in column 5 are extracted from the EHB in-phase oscillation frequency (2) of Ref. [13] which is valid for arbitrary $q$ and $d$ values.

Thus, we may conclude that the 2D point-dipole model reasonably well emulates the in-phase mode of the EHB.

At the present time, only the classical MD simulations generate a direct description of the dynamics of the collective modes. The recent Ref. [9(a)] quantum MC simulations of a strongly coupled bosonic dipole system at zero temperature, however, provide some indirect insight into the collective mode structure at $T=0$. The comparison with the previously obtained QLCA and MD results can be afforded on three different levels: first, the average potential $J\left(\tilde{\Gamma}_D\right)$ in Eq. (18) can be replaced by its $T=0$ equivalent; second, the fitted ground-state energy equation of state formula, as given by Ref. [9(a)] can be employed to find the thermodynamic sound speed; and third, in order to obtain the collective mode dispersion from static structure function data, the Feynman construction, as used by Ref. [9(a)], can be invoked. The first step is made possible by recalling that $(n/2)\int d^2\mathbf{r}\phi_D(r)g(r)=(p^2/a^3)J(r_D)$ is the dipole-dipole interaction energy per particle $E_{\text{int}}(r_D)$; $r_D = r_0/a$ is the effective coupling parameter (defined above) for the zero-temperature 2D dipolar fluid. From the Ref. [9(a)] fitting formula for the ground state energy, we identify the interaction energy as the leading $(nr_0^2)^{3/2}$ term in the series. Introducing the dipole equivalent of the Rydberg energy [9(a)]

$$E_0 = \frac{p^2}{r_0^3} = \frac{\hbar^6}{p^4 m^3} = \frac{\hbar^2}{mr_0^2}, \tag{41}$$

one finds the relation

$$E_{int} = a_1(nr_0^2)^{3/2} E_0 = \frac{p^2}{a^3} J(r_D). \tag{42}$$

Taking $a_1 = 4.536$ from the Ref. [9(a)] fitting formula, one can then calculate

$$J(r_D) = \frac{a_1}{\pi^{3/2}} = 0.8146 \tag{43}$$

as the (apparently independent of $r_D$) value of the integral, in reasonable agreement with its classical equivalent for $\Gamma_D \geq 60$. The QLCA sound velocity

$$s = 1.30\omega_D a \tag{44}$$

then results from Eqs. (17) and (42). This value can be compared with the other data in Table 2; note in particular, it is quite close to the $s = 1.314\omega_D a$ value at $\Gamma_D = 60$.

Addressing the second comparison, the simulation of Ref. [9(a)] reports that the 2D dipolar fluid crystallizes at $nr_0^2 = 290$ (corresponding to $r_D = 30.18$). We therefore select $nr_0^2 =$ 32, 54, 128, 256 ($r_D =$ 10, 14.2, 20, and 28.4) as representative of the strongly coupled fluid-phase for our comparison. These values, when plugged into (40), result in the thermodynamic sound speeds tabulated in Table 3. The thermodynamic sound speeds are higher than the QLCA kinetic sound speed $= 1.3a\omega_D$. This discrepancy decreases from 7.8% to 3% as the coupling parameter increases from $r_D = 10$ to 28.4.

Addressing the third comparison, from Ref. [9a] we calculate the quantum MC sound speed based on Feynman's relation

$$\hbar\omega(q\to 0)=\frac{\hbar^2 q^2}{2mS(q\to 0)}, \tag{45}$$

It appears that the phase velocity extracted from (45) compares less satisfactorily with the thermodynamic sound speed (40) and with the corresponding QLCA value. In this connection, we refer to Figure 3 in Ref. [9(a)] showing the dispersion curves generated from (45) with input of QMC $S(q)$ data (note: the vertical axis in that figure is incorrectly labeled; it should read $mE_k/n\hbar^2$). Choosing the wave number value $q=0.5\sqrt{n}$ in the acoustic domain and (roughly) reading the $mE_k/n\hbar^2$ values off the Fig, 3 $nr_0^2 =$ 32, 64, 128, 256 fluid phase curves, results in the quantum MC sound speeds tabulated in Table 3. We find that the thermodynamic sound speed is 22-34% lower than the QMC sound speed. While it is true that the Feynman excitation spectrum (45) constitutes an upper bound to the actual collective mode dispersion, it is also the case that (45) should reasonably well describe the dispersion in the acoustic regime, especially for zero-temperature bosons. This would imply that the sizable discrepancies could be due to possible inaccuracies in the input $S(q\to 0)$ data in [9(a)]. In any case, the resolution of this issue is not in the purview of the present work.

    We can also compare the classical and quantum QLCA sound speeds as given in Tables 2 and 3, and we observe the marked closeness of the two.

## VI. STLS DESCRIPTION

We have already stated the philosophy that leads us to pursue the STLS calculation, in addition to the QLC analysis already carried out, even though the comparison with various simulation results convincingly demonstrates the reliability of the QLCA scheme. What makes the STLS method attractive is that it has both a classical and a quantum formulation and that the latter is derived from first principles, without recourse to the heuristic arguments exploited in the quantum generalization of the QLCA. Thus, what we are interested in here is less the actual value of the sound velocity, as predicted by the STLS scheme, but rather seeing whether the calculation corroborates the conclusion we have arrived at through the QLCA analysis, namely that in the $q \to 0$ limit there is no difference between the classical and quantum architectures of the point dipole system's longitudinal collective mode.

First, we adapt the STLS kinetic equation approximation scheme [15] to the calculation of the density response function and long-wavelength dispersion for the 2D dipolar fluid in the high-temperature classical domain. The starting point for the calculation is the Fourier-transformed linearized kinetic equation for the perturbed one-particle distribution function $f^{(1)}(\mathbf{v},\mathbf{r},\omega)$, which, in the presence of a weak external dipole potential energy $\Phi_D^{ext}(\mathbf{q},\omega)$, is given by

$$[\omega - \mathbf{q}\cdot\mathbf{v}] f^{(1)}(\mathbf{v},\mathbf{q},\omega) + \frac{1}{m}\mathbf{q}\cdot\frac{\partial f_0^{(1)}(v)}{\partial \mathbf{v}} \Phi_D^{ext}(\mathbf{q},\omega)$$

$$= \frac{i}{m}\frac{\partial}{\partial \mathbf{v}}\cdot\int d^2\mathbf{r}\exp(-i\mathbf{q}\cdot\mathbf{r})\int d^2\mathbf{r}'\int d^2\mathbf{v}' f^{(2)}(\mathbf{v},\mathbf{r};\mathbf{v}',\mathbf{r}';\omega)\nabla\frac{p^2}{|\mathbf{r}-\mathbf{r}'|^3} ; (46)$$

$f^{(2)}(\mathbf{v},\mathbf{r};\mathbf{v}',\mathbf{r}';\omega)$ is the perturbed two-particle velocity distribution function; $f_0^{(1)}(v) = n(\beta m/2\pi)\exp(-\beta m v^2/2)$ is the Maxwellian distribution normalized to the average 2D density $n$; $n(\mathbf{q},\omega) = \int d^2\mathbf{v} f^{(1)}(\mathbf{q},\omega)$ is the average density response. Introducing the equilibrium pair distribution function $g(|\mathbf{r}-\mathbf{r}'|)$, we then make use of the STLS closure hypothesis

$$f^{(2)}(\mathbf{v},\mathbf{r};\mathbf{v}',\mathbf{r}';\omega) = [f_0^{(1)}(v)f^{(1)}(\mathbf{v}',\mathbf{r}',\omega) + f^{(1)}(\mathbf{v},\mathbf{r},\omega)f_0^{(1)}(v')]g(|\mathbf{r}-\mathbf{r}'|) \quad (47)$$

which, when substituted into (46) gives

$$[\omega - \mathbf{q}\cdot\mathbf{v}]f^{(1)}(\mathbf{v},\mathbf{q},\omega) + \frac{1}{m}\mathbf{q}\cdot\frac{\partial f_0^{(1)}(v)}{\partial \mathbf{v}}\Phi_D^{ext}(\mathbf{q},\omega)$$

$$= -\frac{3ip^2}{m}\frac{\partial f_0^{(1)}(v)}{\partial \mathbf{v}}\cdot n(\mathbf{q},\omega)\int d^2\mathbf{R}\frac{\mathbf{R}}{R^5}g(R)\exp(-i\mathbf{q}\cdot\mathbf{R}), \quad (48)$$

where $\mathbf{R} = \mathbf{r}-\mathbf{r}'$. Solving for $f^{(1)}(\mathbf{v},\mathbf{q},\omega)$ and taking the density moment, one readily obtains

$$n(\mathbf{q},\omega) = \frac{\chi_0^V(\mathbf{q},\omega)}{1 - \Lambda(\mathbf{q})\chi_0^V(\mathbf{q},\omega)}\Phi_D^{ext}(\mathbf{q},\omega); \quad (49)$$

$$\Lambda(\mathbf{q}) = \frac{3p^2}{q^2}\int d^2\mathbf{r}\frac{\mathbf{q}\cdot\mathbf{r}}{r^5}g(r)\exp(-i\mathbf{q}\cdot\mathbf{r}) = \frac{6\pi p^2}{q}\int_0^\infty dr\frac{1}{r^3}g(r)J_1(qr); \quad (50)$$

Note that

$$\Lambda(q\to 0) = 3\pi p^2\int_0^\infty dr\frac{1}{r^2}g(r) = \frac{3}{2}\pi a^2 <\phi_D(r)> \quad (51)$$

One can observe the natural emergence of the Vlasov function

$$\chi_0^V(\mathbf{q},\omega) = -\frac{1}{m}\int d^2\mathbf{v}\frac{\mathbf{q}\cdot\partial f_0^{(1)}(v)/\partial \mathbf{v}}{\omega - \mathbf{q}\cdot\mathbf{v}} \quad (52)$$

in the formalism. Comparing (49) and the constitutive relation (9), we obtain

$$\chi(\mathbf{q},\omega) = \frac{\chi_0^V(\mathbf{q},\omega)}{1-\Lambda(\mathbf{q})\chi_0^V(\mathbf{q},\omega)}. \tag{53}$$

When compared with the QLCA, the replacement of $\Psi(\mathbf{q})$ with $\Lambda(\mathbf{q})$ is the hallmark of the STLS approach. Then following the same pattern of reasoning, one finds, similarly to (16)

$$\omega^2(q \to 0) = 3\frac{\pi n p^2 q^2}{m}\int_0^\infty dr \frac{1}{r^2} g(r) = \frac{3}{2} J(\Gamma_D)\omega_D^2 a^2 q^2. \tag{54}$$

or

$$s = \omega_D a \sqrt{\frac{3}{2} J(\Gamma_d)} \tag{55}$$

We therefore recover the acoustic phase velocity (17) with the QLCA $K = 33/32$ value therein replaced by the STLS $K = 3/4$ value.

We turn now to the STLS description of the collective mode in the quantum domain. The analysis is facilitated by adapting Niklasson's quantum kinetic equation for the 3D electron fluid [19] to the 2D dipolar bosonic fluid at arbitrary temperature. Referring to the definitions of the distribution functions provided in Ref. [19], the starting point for our calculation is the linearized kinetic equation for the perturbed one-particle Wigner distribution function (WDF), $f_\mathbf{k}^{(1)}(\mathbf{q},\omega)$, which, in the presence of the weak external dipole potential energy $\Phi_D^{ext}(\mathbf{q},\omega)$, is given by

$$[\hbar\omega-(\hbar^2/m)\mathbf{k}\cdot\mathbf{q}]f_{\mathbf{k}}^{(1)}(\mathbf{q},\omega)-\frac{1}{A}\left[n_{\mathbf{k}-\mathbf{q}/2}-n_{\mathbf{k}+\mathbf{q}/2}\right]\Phi_D^{ext}(\mathbf{q},\omega)$$

$$=\frac{1}{A}\int d^2\mathbf{r}\phi_D(r)\sum_{\mathbf{q}',\mathbf{k}'}\exp(-\mathbf{q}'\cdot\mathbf{r})\left[f_{\mathbf{k}-\mathbf{q}'/2,\mathbf{k}'}^{(2)}(\mathbf{q}-\mathbf{q}',\mathbf{q}';\omega)-f_{\mathbf{k}+\mathbf{q}'/2,\mathbf{k}'}^{(2)}(\mathbf{q}-\mathbf{q}',\mathbf{q}';\omega)\right];$$

(56)

$f_{\mathbf{k}-\mathbf{q}'/2,\mathbf{k}'}^{(2)}(\mathbf{q}-\mathbf{q}',\mathbf{q}';\omega)$ is the perturbed two-particle WDF; $n_{\mathbf{k}}$ is the momentum distribution function for particles with energy spectrum $\varepsilon_{\mathbf{k}}=\hbar^2k^2/(2m)$; $n=(1/A)\sum_{\mathbf{k}}n_{\mathbf{k}}=N/A$ is the average 2D density; $n(\mathbf{q},\omega)=\sum_{\mathbf{k}}f_{\mathbf{k}}^{(1)}(\mathbf{q},\omega)$ is the perturbed density response to $\Phi_D^{ext}(\mathbf{q},\omega)$. Introducing the pair distribution function $g(r)$ with Fourier transform $g(q)$, we now invoke the linearized STLS hypothesis [17]

$$f_{\mathbf{k}\pm\mathbf{q}'/2,\mathbf{k}'}^{(2)}(\mathbf{q}-\mathbf{q}',\mathbf{q}';\omega)=\frac{1}{A}\left[n_{\mathbf{k}\pm\mathbf{q}'/2}f_{\mathbf{k}'}^{(1)}(\mathbf{q},\omega)g(|\mathbf{q}-\mathbf{q}'|)+n_{\mathbf{k}'}f_{\mathbf{k}\pm\mathbf{q}'/2}^{(1)}(\mathbf{q},\omega)g(q')\right]$$

(57)

which, when substituted into Eq. (56), gives

$$f_{\mathbf{k}}^{(1)}(\mathbf{q},\omega)=\frac{1}{A}\left[\frac{n_{\mathbf{k}-\mathbf{q}/2}-n_{\mathbf{k}+\mathbf{q}/2}}{\hbar\omega-(\hbar^2/m)\mathbf{k}\cdot\mathbf{q}}\right]\Phi_D^{ext}(\mathbf{q},\omega)$$

$$+n(\mathbf{q},\omega)\int d^2r\phi_D(r)\frac{1}{A^2}\sum_{\mathbf{q}'}\left[\frac{n_{\mathbf{k}-\mathbf{q}'/2}-n_{\mathbf{k}+\mathbf{q}'/2}}{\hbar\omega-(\hbar^2/m)\mathbf{k}\cdot\mathbf{q}}\right]g(|\mathbf{q}-\mathbf{q}'|)\exp(-i\mathbf{q}'\cdot\mathbf{r})\quad(58)$$

Upon taking the density moment of (58) and comparing the result with constitutive relation (9), one readily obtains the quantum STLS density response function for the 2D dipolar liquid:

$$\chi(\mathbf{q},\omega) = \frac{\chi_0^L(\mathbf{q},\omega)}{1 - \int d^2r \phi_D(r) \frac{1}{A} \sum_{\mathbf{q}'} g(|\mathbf{q}-\mathbf{q}'|) \chi_0^L(\mathbf{q},\mathbf{q}',\omega) \exp(-i\mathbf{q}'\cdot\mathbf{r})}. \tag{59}$$

Note the natural emergence of the inhomogeneous Lindhard function [16, 17]

$$\chi_0^L(\mathbf{q},\mathbf{q}',\omega) = \frac{1}{A} \sum_{\mathbf{k}} \left[ \frac{n_{\mathbf{k}-\mathbf{q}'/2} - n_{\mathbf{k}+\mathbf{q}'/2}}{\hbar\omega - (\hbar^2/m)\mathbf{k}\cdot\mathbf{q}} \right]; \tag{60}$$

$\chi_0^L(\mathbf{q},\omega) = \chi_0^L(\mathbf{q},\mathbf{q},\omega)$. At long wavelengths and in the strong coupling regime,

$$\chi_0^L(\mathbf{q}\to 0,\mathbf{q}',\omega) = n(\mathbf{q}\cdot\mathbf{q}')/(m\omega^2) \tag{61}$$

Consequently,

$$\lim_{\mathbf{q}\to 0} \int d^2\mathbf{r} \phi_D(r) \frac{1}{A} \sum_{\mathbf{q}'} g(|\mathbf{q}-\mathbf{q}'|) \chi_0^L(\mathbf{q},\mathbf{q}',\omega) \exp(-i\mathbf{q}'\cdot\mathbf{r})$$

$$= \frac{n}{m\omega^2} \int d^2\mathbf{r} \phi_D(r) \frac{1}{A} \sum_{\mathbf{q}'} (\mathbf{q}\cdot\mathbf{q}') \exp(-i\mathbf{q}'\cdot\mathbf{r}) \int d^2\mathbf{r}' g(r') \exp[-i(\mathbf{q}-\mathbf{q}')\cdot\mathbf{r}']$$

$$= \frac{in}{m\omega^2} \int d^2\mathbf{r} \phi_D(r) \int d^2\mathbf{r}' g(r') \exp(-i\mathbf{q}\cdot\mathbf{r}')(\mathbf{q}\cdot\nabla) \frac{1}{A} \sum_{\mathbf{q}'} \exp[-i\mathbf{q}'\cdot(\mathbf{r}-\mathbf{r}')]$$

$$= \frac{in}{m\omega^2} \int d^2\mathbf{r} \phi_D(r)(\mathbf{q}\cdot\nabla)[g(r)\exp(-i\mathbf{q}\cdot\mathbf{r})]$$

$$\approx \frac{n}{m\omega^2} q_\mu q_\nu \int d^2\mathbf{r} \phi_D(r) \partial_\mu [g(r) r_\nu]$$

$$= -\frac{n}{m\omega^2} q_\mu q_\nu \int d^2\mathbf{r} g(r) r_\nu \partial_\mu \phi_D(r)$$

$$= \frac{3nq^2}{2m\omega^2} \int d^2\mathbf{r} \phi_D(r) g(r) = \frac{3\pi nq^2 p^2}{m\omega^2} \int_0^\infty dr \frac{1}{r^2} g(r) \tag{62}$$

Eqs. (59) and (62) then give

$$\chi(\mathbf{q}\to 0,\omega) = \frac{\chi_0^L(\mathbf{q}\to 0,\omega)}{1-\Lambda(q\to 0)\chi_0^L(\mathbf{q}\to 0,\omega)}; \tag{63}$$

$\Lambda(q\to 0)$ is given by Eq. (51).

The resulting long-wavelength collective mode frequency

$$\omega^2(\mathbf{q}\to 0) = 3\frac{\pi n q^2 p^2}{m}\int_0^\infty dr \frac{1}{r^2}g(r) = \frac{3}{2}J(\tilde{\Gamma}_D)\omega_D^2 a^2 q^2, \tag{64}$$

which follows from (64), is seen to be identical to its classical counterpart (54). Thus, we arrive at the conclusion that in the long-wavelength limit the classical and quantum dispersion relations are identical.

Digressing now on the question of the accuracy of the STLS theory, we see that the classical STLS sound speeds calculated from (55) are 9.3–10% lower than the thermodynamic sound speeds in Table 1, while at zero temperature, comparing the kinetic sound speed extracted from (64) with the thermodynamic sound speed (40) and using the Eq. (43) $J(r_D) = 0.8146$ value at coupling strength $nr_0^2 = 256$ ($r_D = 28.4$), we find that the STLS kinetic sound speed is 15.6% lower than the thermodynamic sound speed. Thus, we are lead to conclude that, of the two theoretical approaches followed in this paper, the QLCA is indeed the superior one.

## VII. CONCLUSIONS

In this paper we have addressed the question of the long-wavelength behavior of the density oscillation mode in a 2D system of interacting dipoles. Our main observation has been that the system does not permit an RPA-type approximation and correlations have to be accounted for from the outset. We have used the QLCA formalism in conjunction with classical MD simulations to determine the mode dispersion. We have argued that, in the domain of interest, the long-wavelength collective mode behavior in the classical system is not different from that in the quantum systems. In addition to the explicit QLCA results, this contention has been corroborated through STLS calculations pertaining both to the classical and quantum domains. We have rigorously shown that the mode behavior is acoustic and that the suggested [5] $\omega(q \to 0) \propto q^{3/2}$ is erroneous. Our principal objective has been to determine the sound velocity associated with this acoustic mode. The results of the ensuing calculations cover the entire classical and quantum domain, all the way down to zero temperature.

Since the point-dipole system is, in fact, assumed to be a faithful representation of the excitonic (dipolar) phase of the electron-hole bilayer system, we have sought to establish the parallelism between the results of the present work and those of a similar investigation pertaining to the EHB [13]. The acoustic velocities extracted from Eqs. (16), (54), and (64) exhibit precisely the same dependence on $d$ and $g(r)$ as the EHB QLCA in-phase velocity (1) in the strong coupling $d \to 0$ limit. Moreover, the architectures of the acoustic velocities of EHB Eq. (1) and 2D dipole system Eq. (17) are identical in every respect. In establishing this equivalence, the role of strong correlations in the EHB is crucial: in the RPA description of the EHB the requisite linear dependence of the sound speed on the layer separation $d$ is absent.

We have also focused on comparing our results with those of recent quantum MC simulations of a zero-temperature bosonic dipole system [9(a)]. We have been able to extend the QLCA calculations to this domain, making use of the ground state energy data obtained from [9(a)], and have found good agreement with the classical sound velocity values. The agreement is less impressive with the sound velocity generated through the Feynman formalism from the static structure function data of [9(a)]; it is not clear whether this disagreement is due to the lack of precision in these QMC data or to some other reason.

We have been able to relate the derived values of the sound speed to the thermodynamic sound speed obtained from classical MD [22] and quantum MC [9(a)] generated equations of state. We have also compared the sound velocities obtained in the strongly coupled liquid phase with the phonon phase velocities in the 2D dipole and EHB classical lattices, obtained by routine lattice sum calculations. All these different approaches converge into a coherent physical picture.

The possible influence of the existing bosonic dipole condensate on the collective mode structure has been ignored. The QMC results show that the condensate fraction does not exceed a few percent either in the 2D dipole system [9(a)] or in the EHB system [3]; thus, while it would be of great interest to address this issue, the quantitative difference from the present results is not expected to be significant.

.**ACKNOWLEDGEMENTS**

This material was based upon work supported by the National Science Foundation under Grants No. PHY-0514618 and PHY-0514619, and by the Hungarian Fund for Scientific Research and the Hungarian Academy of Sciences, OTKA-T-48389, OTKA-IN-69892, MTA-NSF-102, OTKA-PD-04999.